\documentstyle[prl,aps,multicol,psfig]{revtex}
\begin{document}
\draft \wideabs{
\title{Existence of orbital polarons in ferromagnetic insulating
La$_{1-x}$Sr$_x$MnO$_{3}$~(0.11$<x<$0.14) evidenced  by giant
phonon softening}
\author{K.-Y. Choi,$^1$ P. Lemmens,$^{1,2}$ G. G\"{u}ntherodt,$^1$
Yu. G. Pashkevich,$^3$  V. P. Gnezdilov,$^4$ P. Reutler,$^{5,6}$
\\L. Pinsard-Gaudart, $^6$ B. B\"{u}chner,$^5$ and   A. Revcolevschi
$^{6}$ }
\address{$^1$ 2. Physikalisches Institut, RWTH Aachen, 52056
Aachen, Germany \\ $^2$ Max Planck Institute for Solid State
Research, D-70569 Stuttgart, Germany
\\ $^3$ A. A. Galkin Donetsk Phystech NASU, 83114 Donetsk, Ukraine \\
$^4$ B. I. Verkin Inst. for Low Temp. Physics NASU, 61164 Kharkov,
Ukraine\\$^5$ Institute for Solid State Research, IFW Dresden,
D-01171 Dresden, Germany \\ $^6$ Laboratoire de Physico-Chimie,
Universit\'{e} Paris-Sud, 91405 Orsay, France}
\date{\today}
\maketitle

\begin{abstract}
We present an inelastic light scattering study of single
crystalline (La$_{1-y}$Pr$_y$)$_{1-x}$Sr$_{x}$MnO$_3$ ($0\leq
x\leq0.14$,\,\,$y=0$ and $x=1/8$,\,\,$0\leq y\leq0.5$). A giant
softening up to 20 - 30~cm$^{-1}$ of the Mn-O breathing mode has
been observed only for the ferromagnetic insulating (FMI) samples
($0.11\leq x \leq 0.14$) upon cooling below the Curie temperature.
With increasing Pr-doping the giant softening is gradually
suppressed. This is attributed to a coupling of the breathing mode
to orbital polarons which are present in the FMI phase.
\end{abstract}

 }

\narrowtext

Orbital degrees of freedom play a crucial role in understanding
the metal-insulator transition, the colossal magnetoresistance
effect, and the phase diagram of manganites \cite{Tokura,BrinkR}.
The unusual coexistence of ferromagnetism and insulating behavior
in lightly doped manganites La$_{1-x}$Sr$_x$MnO$_3$
($x=0.11-0.18$) \cite{Dabrowski99,Urushibara95,Cox01,klingerer}
highlights the significance of orbital degrees of freedom beyond
the double exchange (DE) mechanism. Despite intensive
investigations  there is, up to now, no consensus on the exact
form of the orbital ordering in this doping range.

In this Letter we report on Raman scattering measurements
performed for the first time on single crystalline samples of
(La$_{1-y}$Pr$_y$)$_{1-x}$Sr$_{x}$MnO$_3$ ($0\leq
x\leq0.14$,\,\,$y=0$ and $x=1/8$,\,\,$0\leq y\leq0.5$). Raman
spectroscopy can shed light on the coupling of lattice and orbital
degrees of freedom as optic phonons are sensitive to changes of
the orbital state via electron-phonon coupling. We observe a giant
and continuous softening by 20 - 30~cm$^{-1}$ of the 610-cm$^{-1}$
Mn-O breathing mode only in the ferromagnetic insulating (FMI)
samples ($0.11\leq x \leq 0.14$) upon cooling below the Curie
temperature. The giant softening originating from an
orbital-phonon coupling gives evidence for the presence of orbital
polarons in the FMI phase. The softening persisting to the lowest
temperature is due to hole-orbital coupling leading to the
instability of the orbital polaron state.

The phase diagram of our lightly doped single crystalline samples
of La$_{1-x}$Sr$_x$MnO$_3$ ~\cite{klingerer} is displayed in
Fig.~1. Upon cooling samples with $0.11\leq x\leq0.14$ show a
cooperative Jahn-Teller (JT) distortion at T$_{JT}$. Below
T$_{JT}$ successive phase transitions occur from a paramagnetic
insulating (PI) to a ferromagnetic metallic (FMM) state at T$_C$
and from a FMM to a FMI phase at T$_{CO}$. Between T$_{JT}$ and
T$_C$ an antiferro-orbital order which is similar to that of
LaMnO$_3$ is present \cite{Geck03}. Between T$_{C}$ and T$_{CO}$
the antiferro-orbital order is strongly suppressed and metallic
resistivity is observed due to DE \cite{Dabrowski99}. In the FMI
phase neutron and x-ray studies \cite{niemoeller,Yama96,Yama00}
show a superstructure interpreted in terms of long-range charge
order. An earlier resonant x-ray scattering (RXS) study
\cite{Endo99} questioned this interpretation and instead suggested
the presence of a new type of orbital ordering. A recent RXS study
\cite{Geck03} substantiates this rearrangement of the orbital
ordering at T$_{CO}$ related to an enhancement of the
ferromagnetic superexchange (SE) with respect to the DE. This
orbital ordering also leads to an upturn of the resistivity and a
step-like increase of the magnetization upon cooling
\cite{Nojiri99}. Concerning the orbital/charge ordered state of
the FMI phase there exists evidence for a periodic structural
modulation with alternating hole-poor and hole-rich planes along
the $c$ axis \cite{Yama96,Yama00,Mizokawa1}.

(La$_{1-y}$Pr$_y$)$_{1-x}$Sr$_x$MnO$_3$ single crystals were grown
using the floating zone  method \cite{Anane}. The high quality of
our single crystals is confirmed by the reported results of x-ray
diffraction \cite{niemoeller},  magnetization,  specific heat
\cite{klingerer}, and RXS~\cite{Geck03}. Raman scattering
measurements were carried out in a quasi-backscattering geometry
with the excitation line $\lambda= 514.5$ nm of an Ar$^{+}$ laser.
The laser power of 7 mW was focused to a 0.1 mm diameter spot on
the sample surface. All spectra were analyzed by a DILOR-XY
spectrometer and a nitrogen cooled CCD detector.

Figure~2 displays  the doping dependence of Raman spectra  of
La$_{1-x}$Sr$_x$MnO$_3$ ($x$=0, 0.06, 0.09, 0.11, 0.125, and 0.14)
at room temperature. For the undoped sample, 15 sharp and intense
peaks are observed which are part of 24 symmetry-allowed modes in
the Pnma space group of the perovskite structure \cite{Iliev98}.
The modes below 330 cm$^{-1}$ arise from vibrations of (La/Sr)
cations and rotations of the MnO$_6$ octahedra. The modes above
400 cm$^{-1}$ are related to bending and stretching vibrations of
the octahedra. With increasing doping $x$ the Raman scattering
intensity of the phonon modes is strongly reduced and their
damping increases because of the weakening of the static JT
distortion as well as the screening effect by charge carriers. As
a result, in the FMI phase only three modes are observable: the
out-of-phase rotational mode around 250 cm$^{-1}$, the JT mode
around 490 cm$^{-1}$, and the breathing mode around 610 cm$^{-1}$
(see Fig.~2 for $x>0.10$).

The temperature dependence of the Raman spectra of two
representative samples ($x=0.06$ and 0.125) is shown in Fig.~3.
Raman spectra of the canted antiferromagnetic (CAF) samples (shown
here at $x=0.06$) show no drastic change except a slight shift and
sharpening of the phonon modes with decreasing temperature. In
contrast, all FMI samples (shown here at $x=0.125$) show the
appearance of new modes and the development of fine structures
upon cooling below T$_{CO}$. The sharpness of the observed phonon
modes below T$_{CO}$ rules out the possibility of
chemical/structural disorder. Rather, these fine structures should
be ascribed to {\it activated} modes which are not Raman-active in
the PI phase and to zone-folded modes caused by the superstructure
related to charge ordering or the discussed stripe formation of
holes. Thus, our results signal a static long-range charge/orbital
ordering.

The most distinct features of this study are observed in the
temperature dependence of the Mn-O bond modes. As shown in the
left panel of Fig. 4, all CAF samples show a similar behavior:
with decreasing temperature the JT mode at about 490 cm$^{-1}$
first hardens by 5-6 cm$^{-1}$ up to T$_N$ and then softens by 3
cm$^{-1}$. Error bars of the fitted data are of the size of the
symbols. The breathing mode at about 610 cm$^{-1}$ shows a
softening by 5-8 cm$^{-1}$ below T$_N$ following a slight
hardening. With increasing doping $x$ through the CAF/FMI phase
boundary the moderate softening below T$_N$ turns over to a more
pronounced one in the FMI phase. In the right panel of Fig.~4 the
JT mode for $x=0.11$ softens by 3 cm$^{-1}$ below T$_C$ while for
$x=0.125$ and 0.14 by 6-7 cm$^{-1}$. In contrast, the breathing
mode undergoes a giant softening by 20 cm$^{-1}$ for $x=0.11$ and
30 cm$^{-1}$ for $x=0.125$ and $x=0.14$. It is worth to note that
the small softening of the JT mode is contrasted by the giant
softening of the breathing mode. Furthermore,  the softening does
not exhibit any saturation at very low temperatures while its
slope becomes less steep.

In the following we will focus on the giant softening of the
breathing mode seen in the FMI samples and discuss its possible
origins. A detailed analysis of phonon modes will be given
elsewhere~\cite{Choi}.

Firstly, we will consider a change of lattice symmetry and
parameters as a possible reason. In contrast to the very lightly
doped CAF samples an appreciable change of lattice parameters
takes place in the FMI samples between T$_{C}$ and T$_{CO}$
because of the suppression of the cooperative JT distortion
\cite{Dabrowski99}. In addition, the crystal symmetry is reduced
to triclinic below T$_{CO}$. Thus, we might attribute the observed
giant softening to lattice anomalies. However, the lattice
contribution to the frequency shift estimated by the Gr{\"u}neisen
law for a nearly cubic lattice, $\Delta\omega/\omega \propto
\Delta V/V$ cannot exceed more than a few cm$^{-1}$. This {\it
expected} small effect is drastically contrasted by the {\it
observed} giant softening. Furthermore, the lattice anomalies
cannot give an account of the drastically different behavior
between the JT and the breathing modes. As will be discussed
below, Pr-doping on La$_{7/8}$Sr$_{1/8}$MnO$_3$ leads to the
suppression of the giant softening (see Fig.~5(a)) in spite of
enhanced lattice modulations~\cite{Pascal}. This necessitates
other mechanisms beyond a minor lattice contribution.

Secondly, magnetic ordering may lead to phonon anomalies
proportional to the square of the magnetic moment of the magnetic
ions~\cite{BaltHel}. Apparently, the onset of the softening starts
around the magnetic ordering temperature. Indeed, this spin-phonon
mechanism can explain the softening of the corresponding modes by
8 cm$^{-1}$ in LaMnO$_3$ below T$_N$ ~\cite{Gran98,Gran99}. All
CAF samples which have the same orbital and crystal structure show
a similar anomaly. This provides evidence for a magnetic origin of
the softening in the CAF samples. In the FMI samples the softening
is strongly enhanced. At first glance, this seems to be related to
an increase of the magnetic ordered moments. However, with
increasing doping $x$ the softening does not scale with the
magnetization~\cite{Dabrowski99}. Furthermore, it would be
difficult to understand why the spin degree of freedom couples
selectively to the breathing mode  and not to the JT mode.

To get more insight into this problem we measured the Raman
spectra in a field of 3 T for $x=0.125$. As Fig.~5(b) shows, we
observe several anomalous features: (i) a shift of the onset of
the softening to higher temperature, (ii) an enhanced softening by
4-5 cm$^{-1}$ between 110 K and 240 K compared to 0 T, and (iii)
the saturation (and even a minor reduction) of the softening below
50 K. Note here that 3 T is strong enough to saturate the
magnetization and to enhance its absolute value by a factor of
seven compared to 0.02 T \cite{klingerer}. The first two features
in the high-temperature region are qualitatively consistent with
field-enhanced magnetic moments. However, the third one in the
low-temperature region is opposite to what is expected. Another
objection can be raised by the Pr-doping effect. Doping with Pr
ions has  little influence on the magnetization while stabilizing
JT distortions concomitantly with the corresponding orbital
state~\cite{Pascal}. Thus, the magnetic mechanism cannot explain
the strong reduction of the giant softening with increasing
Pr-content.  Therefore, the main mechanism is not of magnetic and
lattice origins.

Finally, a more relevant mechanism for a large phonon softening
can rely on the coupling of phonons to orbital degrees of
freedom~\cite{BrinkR,Ishi97,Bala,Brink}. Through the
metal-insulator transition (MIT) around $x=0.125$ the
rearrangement in the orbital sector occurs from the
antiferro-orbital ordered state for T$_{C}<$ T $<$ T$_{JT}$ into a
new orbital ordered state for T $<$ T$_{CO}$~\cite{Geck03}. In the
FMM phase (see Fig.~1) inelastic neutron scattering shows large
orbital fluctuations~\cite{Moussa03}. Ultrasonic measurements
exhibit a pronounced softening of the transverse
($C_{11}-C_{12}$)/2 mode which is caused by charge fluctuations
due to enhanced DE ~\cite{Hazama00}. Below T$_{CO}$ holes become
localized and a new orbital state is stabilized. Noticeably, the
development of this state is closely related to the mobility of
holes. This indicates that a new orbital state evolves from the
antiferro-orbital ordered state by polarizing orbitals toward
their neighboring localized hole-rich sites~\cite{Kilian99}.
Orbital polarons in turn are charge carriers dressed by orbitals
which are directed toward the neighboring holes.

As emphasized above, the giant softening is seen only in the
breathing mode within the FMM phase. The displacement pattern of
the breathing mode is directly coupled to the formation of orbital
polarons ~\cite{Kilian99,Mizokawa2}. Thus, one can expect a
pronounced frequency shift of the breathing mode through the MIT.
In this orbital-phonon coupling mechanism the renormalization of
the breathing mode $\Omega_{br}=615$ cm$^{-1}$ will be determined
by (i) the orbital-phonon matrix element $M$ and (ii) the change
of energy $\delta\Delta$ in an electronic state  through the MIT.
In second order perturbation theory the renormalized energy
$\delta\Omega_{br}$ can be estimated as follows \cite{GG}
\begin{equation}
\delta\Omega_{br}=-\frac{|M|^{2}}{2m\Omega_{br}}
\delta(\frac{1}{\Delta-\hbar\Omega_{br}}+\frac{1}{\Delta+\hbar\Omega_{br}}),
\end{equation}
where $m$ is the atomic mass of Mn. Supposing that the MIT is
initiated by the formation of orbital polarons, $\delta\Delta$ can
be approximated by the binding energy of orbital polarons
$\Delta_{op}=0.6$ eV~\cite{Kilian99}. Further, since an elastic
softening of the ($C_{11}-C_{12}$)/2 mode is closely related to
charge/orbital ordering~\cite{Hazama00},  $|M|^2$ can be evaluated
as $|M|^{2}=\frac{3}{4}a_{0}[\delta
(\frac{C_{11}-C_{12}}{2})]\delta(\Delta)\approx0.22$
(eV/$\AA$)$^2$~\cite{GG} using a lattice constant $a_{0}=5.5\AA$
and $\delta(\frac{C_{11}-C_{12}}{2})\approx 1.4\times10^{11}$
erg/cm$^3$ within the FMM phase~\cite{Hazama00}. In the above
estimate the factor of $1/2$ is multiplied since orbital polarons
are formed mainly in hole-rich planes. Inserting this into Eq.~(1)
one finally obtains $\delta\Omega_{br}\approx -74$ cm$^{-1}$. This
estimated value agrees reasonably with the observed phonon shift
of $-30$ cm$^{-1}$ considering the simple assumption for
$\delta\Delta$ and a full polarization of orbital polarons. This
suggests that the softening of the breathing mode origins from its
intrinsic coupling to these orbital polarons which develop in a
commensurate charge ordered region around $x=1/8$. Our study thus
uncovers the presence of orbital polarons in the FMI phase which
naturally explain the concomitant occurrence of ferromagnetism and
insulating behavior. Regarding the small softening of the JT-mode
it should be ascribed to the boundary between two commensurate
charge-ordered regions and/or the presence of another type of
orbital state in the hole-poor plane.

In this situation, the breathing mode may be considered as a part
of a complex order parameter that exhibits a soft mode behavior.
This softening is expected to saturate below T$_{CO}$. However,
the softening persists even to the lowest temperature. This
implies that the orbital polaron state is not robust with respect
to the hopping of charge carriers, pointing to the remaining
significance of DE and hole-orbital coupling. Our observations are
in a good agreement with the strong incoherent spectral weight in
optical spectroscopy arising from scattering of charge carriers by
orbital fluctuations~\cite{Okimoto97}.

The presence of orbital polarons also accounts for the Pr-doping
and the external magnetic field effect. The chemical pressure
exerted by the small radius of the Pr ions  leads to a
stabilization of local JT distortions competing with orbital
polarons~\cite{Pascal}. With increasing Pr-content orbital
polarons turn into the antiferro-orbitals. This leads to the
suppression of the giant softening and indicates that orbital
polarons can be easily destroyed in the presence of a lattice
distortion. Regarding the field dependence an external field
stabilizes orbital polarons through orbital-spin coupling.
Therefore, one expects a saturation of the breathing mode and a
shift of the onset of the softening to higher temperature as
observed in Fig.~5(b).

In summary, our Raman study on single crystals of the lightly
doped manganites La$_{1-x}$Sr$_x$MnO$_3$  shows a  giant softening
of the Mn-O breathing mode  only in the FMI phase persisting to
the lowest temperature. This result gives strong evidence for
orbital polaron formation on the hole-rich sites as well as the
instability of orbital polarons at the lowest temperatures due to
hole-orbital scattering. Furthermore, the rapid suppression of
orbital polarons by Pr doping suggests that their existence is
strongly conditioned by the motion of holes, magnetic
correlations, and lattice distortions.

We thank J. Geck, R. Klingeler, and C. Bauman  for useful
discussions. The work  was supported by DFG/SPP 1073, INTAS 01-278
and NATO Collaborative Linkage Grant PST.CLG.977766.

Fig. 1 Phase diagram of La$_{1-x}$Sr$_x$MnO$_3$ near the
ferromagnetic insulating phase adapted from Klingeler {\it et
al.}\\

Fig. 2 Doping dependence of Raman spectra of
La$_{1-x}$Sr$_x$MnO$_3$ ($x$=0, 0.06, 0.09, 0.11, 0.125, and 0.14)
at 295 K.\\

Fig. 3 Temperature dependence of Raman spectra of two
representative La$_{1-x}$Sr$_x$MnO$_3$ ($x$= 0.06 and 0.125) at
295, 250, 200, 150, 100, 50 and 5 K.\\

Fig. 4 Temperature dependence of peak position of the 490
cm$^{-1}$ and 610 cm$^{-1}$ Mn-O stretching modes. Arrows indicate
the magnetic ordering temperatures. \\

Fig. 5 (a) Pr-doping effect on
(La$_{1-y}$Pr$_y$)$_{7/8}$Sr$_{1/8}$MnO$_3$ (y=0, 0.1, 0.25 and
0.5). (b) Comparison of temperature dependence of the 610
cm$^{-1}$-mode at 0 T and 3 T for $x=0.125$ and $y=0$. \\

\end{document}